\documentclass[11pt,a4paper]{article}
\usepackage{jcappub}
\usepackage{graphicx}
\usepackage{amsfonts}
\usepackage{amssymb}
\usepackage{cite}

\title{How accurately can we measure the hydrogen $2S\rightarrow 1S$ transition rate from the cosmological data?}

\author[a,b]{Viatcheslav Mukhanov}
\author[c,d]{Jaiseung Kim}
\author[c,d]{Pavel Naselsky}
\author[e,f]{Tiziana Trombetti}
\author[f,g]{Carlo Burigana}
\affiliation[a]{Arnold-Sommerfeld-Center, Department f\"ur Physik, Ludwig-Maximilians-Universit\"at M\"unchen, Theresienstr. 37, D-80333, Munich, Germany}
\affiliation[b]{LPT de l'Ecole Normale Superieure, Chaire Blaise Pascal, 24 rue Lhomond, 75231 Paris cedex, France}
\affiliation[c]{Niels Bohr Institute, Blegdamsvej 17, DK-2100 Copenhagen, Denmark}
\affiliation[d]{Discovery Center, Blegdamsvej 17, DK-2100 Copenhagen, Denmark}
\affiliation[e]{Dipartimento di Fisica, Universit`a La Sapienza, P.le A. Moro 2, I-00185 Roma, Italy}
\affiliation[f]{INAF-IASF Bologna, Via P. Gobetti 101, I-40129, Bologna, Italy}
\affiliation[g]{ Dipartimento di Fisica, Universit`a degli Studi di Ferrara, Via G. Saragat 1, I-44100 Ferrara, Italy}
\emailAdd{mukhanov@physik.lmu.de}
\emailAdd{jkim@nbi.dk}
\emailAdd{naselsky@nbi.dk}
\emailAdd{trombetti@iasfbo.inaf.it}
\emailAdd{burigana@iasfbo.inaf.it}

\date{\today}

\abstract{Recent progress in observational cosmology, and especially the forthcoming  PLANCK mission data, open new directions in so-called precision cosmology. In this paper we illustrate this statement considering the accuracy of cosmological determination of the two-quanta decay rate of 2s hydrogen atom state. We show that  the PLANCK data will allow us to measure this decay rate significantly better than in the laboratory experiments.}
\arxivnumber{}

\begin{document}
\maketitle
\section{Introduction}
Since the year 2000 the modern cosmology entered the stage which can be
characterized as an epoch of \textquotedblleft precision
cosmology\textquotedblright. After Saskatoon, TOCO, BOOMERAMG and MAXIMA-1
data, and then through the WMAP, CBI, ACBAR, ACT and getting closer to the
data release of the PLANCK mission, our knowledge of the Universe becomes
more and more informative
\citep{Saskatoon,MAT,BOOMERANG_flat,MAXIMA-1,CBI:observation,WMAP7:basic_result,
WMAP7:powerspectra,WMAP7:Cosmology,ACBAR2008,ACT2008,ACT_powerspectrum}
. There is no doubt that merging the micro-physics learned on the Large
Hadronic Collider (LHC) with macro-physics discovered in space missions like
WMAP and PLANCK \citep{WMAP7:basic_result,WMAP7:powerspectra},
\citep{Planck_bluebook,Planck_mission_early}, will make the picture of the
evolving Universe more \textquotedblleft colorful\textquotedblright\ and
self-consistent.

In this paper we would like to illustrate the current status and
perspectives of the \textquotedblleft precision cosmology\textquotedblright
, considering a simple question, namely, with which accuracy one can measure
the rate of  the two-photons decay for $2S\rightarrow 1S$ transition in the
hydrogen atom from cosmological data. Note that the process of cosmological
hydrogen recombination crucially depends on this process
\citep{Recombination_Zeldovich1,Recombination_Zeldovich2,Recombination_Peebles,Recombination_impact_Rubino}.

 From Quantum Electrodynamic we know the theoretical value for
corresponding decay rate: $A_{2s1s}=8.227~s^{-1}$ (see
\citep{W2s1s_Goldman,W2s1s_Spitzer,Recombination_Prim, Induced_2s,
Chluba_Bernard}).
 However, although there are no doubts
about this process, there is very little experimental verification of it
because the corresponding experiments are very difficult
\citep{Two_photon_decay_OConnell}. Experimentally, 2S-1S two photon
transition has been measured in
\citep{two_photon_molybdenum,two_photon_atomic,two_photon_gold} for the decay of K-shell vacancy in initially neutral atom
using the photon-photon coincidence technique. In these experiments, the
K-shell vacancy is produced by irradiating the targets
by photons or radioactive isotopes, preferable decaying by nuclear
electron capture . However, all these experiments mainly are
devoted to investigation of the heavy ions, rather then hydrogen atom.

On the other hand for the hydrogen  the two photon decay of $2S$ -level
determines the rate of recombination in the \textquotedblleft middle of
recombination layer\textquotedblright , where the pattern of the CMB
polarization is basically formed, we can hope that the precision
cosmological data could allow us to estimate $A_{2s1s}$-constant with rather
high accuracy. At least, using the CMB data in combination with HST,
BAO,SDSS data sets, or
CMB temperature and polarization along (like at PLANCK experiment), we can
estimate the range of
uncertainties of $A_{2s1s}$, as theoretical, as experimental one, which
could give a significant
impact to most probable value of the cosmological parameters
(the barionic density $\Omega_b$, the Cold Dark Matter density
$\Omega_{\mathrm{DM}}$, the Dark Energy density $\Omega_{DE}$,
the spectral index of the adiabatic perturbations $n_s$ etc). This range
of uncertainties can restrict even
theoretical improvement of the kinetic of the hydrogen  recombination,
which in general is very complex and requires incorporation
of $nS\rightarrow 1S$ and $nd\rightarrow 1S$ transitions for levels $n\gg
1$(see for review \citep{Chluba_Recombination}).

We will use the WMAP 7 TT and TE observational data
\citep{WMAP7:basic_result,WMAP7:powerspectra} and the PLANCK mock data
respectively, and show that the current cosmological data give us:
$A_{2s,1s}\simeq 8_{-1.8}^{+3.85}$ (see Section 4). Given the expected
sensitivity of the PLANCK data, we will show that the estimated accuracy can
be further significantly improved: $8.086\mathrm{s}^{-1}<A_{2s1s}<9.037
\mathrm{s}^{-1}$ and $7.613\mathrm{s}^{-1}<A_{2s1s}<9.505\mathrm{s}^{-1}$ at
1$\sigma $ and 2$\sigma $ level respectively.

\section{ Recombination of cosmological hydrogen}

\label{ionization} After $\mathrm{He}^{4}$ recombination the ionization
history characterized by the free electron fraction $x_{e}$ over redshift $z
$ is described by the following equation %
\citep{Recombination_Zeldovich1,Recombination_Zeldovich2,Recombination_Peebles}%
: 
\[
\frac{dx_{e}}{dz}=\frac{1}{(1+z)H(z)}\,C\,[\alpha _{c}nx_{e}^{2}-\beta
_{c}(1-x_{e})\exp \left( -\frac{B_{1}-B_{2}}{k_{B}T}\right) ],\label{xe_evolution}
\]%
where $H(z)$ is the Hubble expansion rate at a redshift $z$, $n$ is the
number density of atoms, $B_{i}$ is the binding energy of hydrogen in the $i$%
th quantum state, $T$ is the temperature of the cosmic plasma, $\alpha _{c}$
and $\beta _{c}$ are the effective recombination and photo-ionization rate
for the states of a principal quantum number greater than one. The factor $C$
in Eq. \ref{xe_evolution} is given by \citep{Recombination_Peebles}: 
\[
C=\frac{1+KA_{2s1s}\,n_{H}(1-x_{e})}{1+KA_{2s1s}\,n_{H}(1-x_{e})+K\,\beta
_{c}\,n_{H}(1-x_{e})},
\]%
where $A_{2s1s}$ is the two-photon decay rate of the 2s hydrogen state and $%
K=\lambda _{\alpha }^{3}/8\pi H(z)$ with the wavelength of Ly-$\alpha $
photon $\lambda _{\alpha }$. This equation is applicable at the range of
redshifts $800\leq z\leq 1100$, providing initial condition for the next
stage ($z\leq 800$), when the two-photon decay is no longer significant %
\citep{Recombination_Peebles}. Currently, the decay rate $A_{2s1s}$ is
theoretically calculated to be $8.22458\,\mathrm{s}^{-1}$, where the slight
improvement in numerical accuracy is made compared to earlier results %
\citep{W2s1s_Goldman,W2s1s_Spitzer}.

\begin{figure}[tbp]
\centering
\includegraphics[width=0.295\textheight]{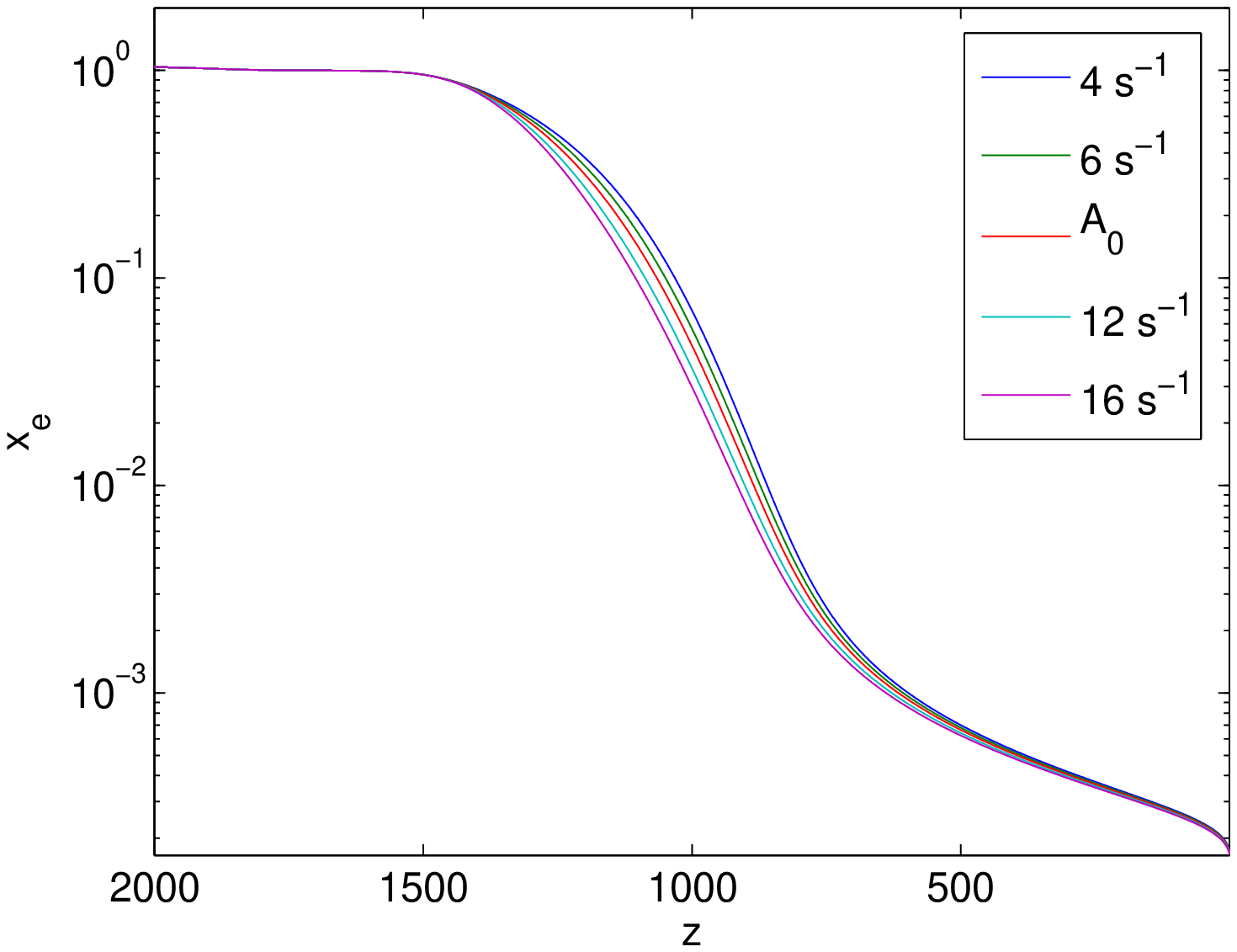}
\includegraphics[width=0.31\textheight]{./figure/fig1b.ps}
\includegraphics[width=0.31\textheight]{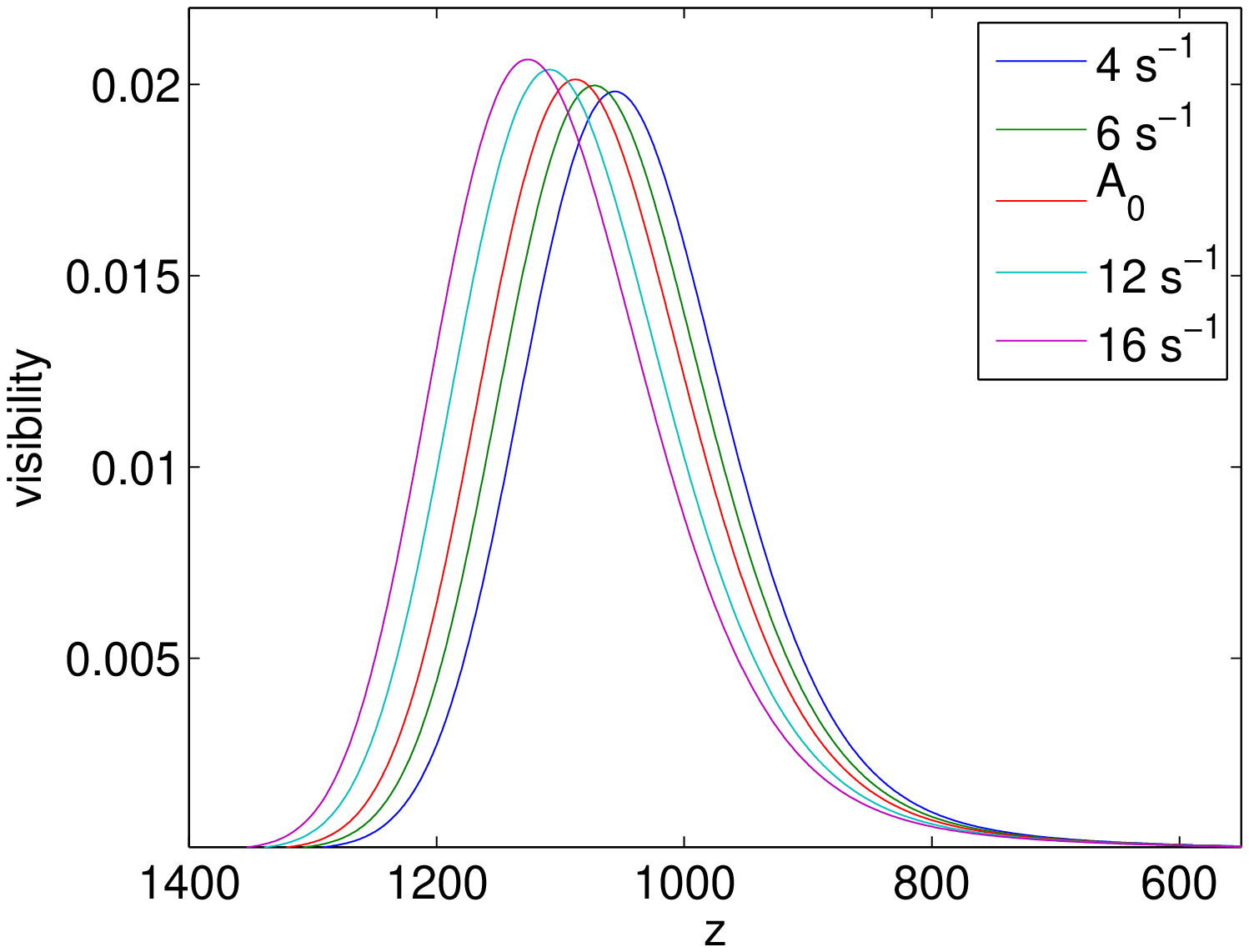}
\caption{Left Top. Ionization history: fraction of free electrons, $x_{e}$, is
plotted over a range of redshift $z$ for various values of $A_{2s1s}$ with $%
A_{0}$ being the theoretical prediction $8.22458\,\mathrm{s}^{-1}$. Right Top.
Relative variations of the ionization fraction $\Delta _{x}=2(x_{e}-%
\overline{x_{e}})/(x_{e}+\overline{x_{e}})$. Here $x_{e}$ corresponds to
current value of $A_{2s1s}$, and $\overline{x_{e}}$ corresponds to
theoretical value of $A_{2s1s}$. 
Bottom. The visibility function, which corresponds to the probability of the last scattering at a redshift $z$, is plotted for various $A_{2s1s}$.
}
\label{xe}
\end{figure}

In Fig. \ref{xe}, we show the ionization history of cosmic plasma for
various decay rate $A_{2s1s}$, where we numerically computed it with the
help of the widely used \texttt{RECFAST} code with a slight modification %
\citep{RECFAST1,RECFAST2,RECFAST3}. 
In the same figure, we plotted the visibility function, which shows the probability of last scattering at a redshift $z$.
As it is seen from Fig. \ref{xe}, the fraction of ionization $x_{e}$ in the
vicinity of $z\simeq 1000$ increases with respect to the standard one, if
the two-photon decay rate $A_{2s1s}$ is lower, and vice versa. 
From Fig. \ref{xe}, we may see that the last scattering occurs at more recent time with slightly wider spread, as the two-photon decay rate $A_{2s1s}$ gets smaller. 
The change of the recombination rate as a function of $A_{2s1s}$
leads to the observational traces in the CMB TT, TE, EE power spectra (refer to \citep{Ionization_history_DM,Ionization_history_CBI,Accelerated_recombination,Probing_Surface,Antimatter_CMB}
for details) and therefore this allows us to determine $A_{2s1s}$ with
rather good accuracy.

\section{CMB anisotropy}

\label{CMB} The whole-sky Stokes parameters of the CMB anisotropy can be
decomposed in terms of spin $0$ and spin $\pm 2$ spherical harmonics: 
\begin{eqnarray*}
\Delta T(\hat{\mathbf{n}}) &=&\sum_{lm}a_{T,lm}\,Y_{lm}(\hat{\mathbf{n}}), \\
Q(\hat{\mathbf{n}})\pm iU(\hat{\mathbf{n}}) &=&\sum_{l,m}-(a_{E,lm}\pm
i\,a_{B,lm})\;{}_{\pm 2}Y_{lm}(\hat{\mathbf{n}}).
\end{eqnarray*}
\begin{figure}[tbp]
\centering
\includegraphics[width=0.34\textheight]{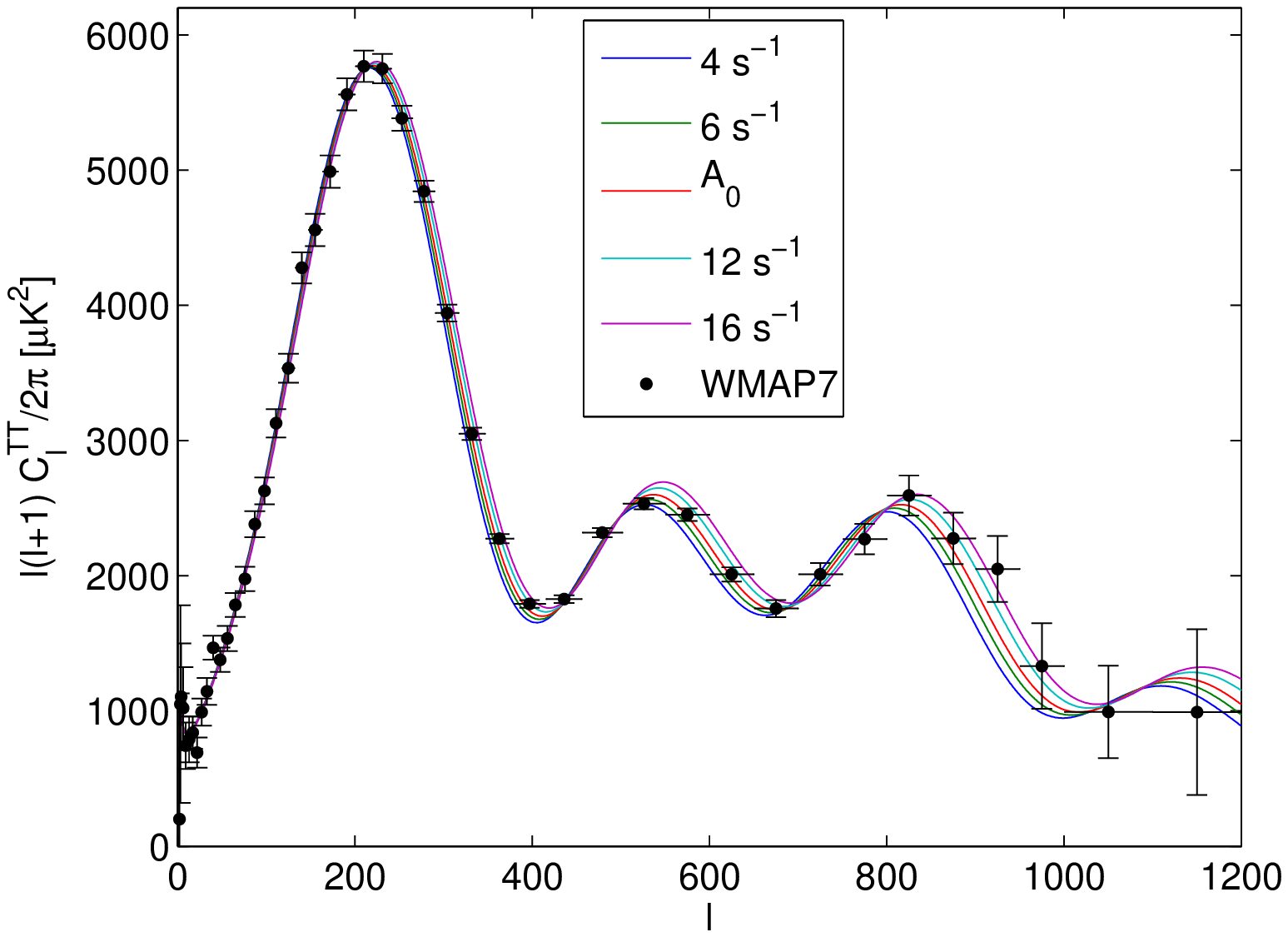}
\includegraphics[width=0.34\textheight]{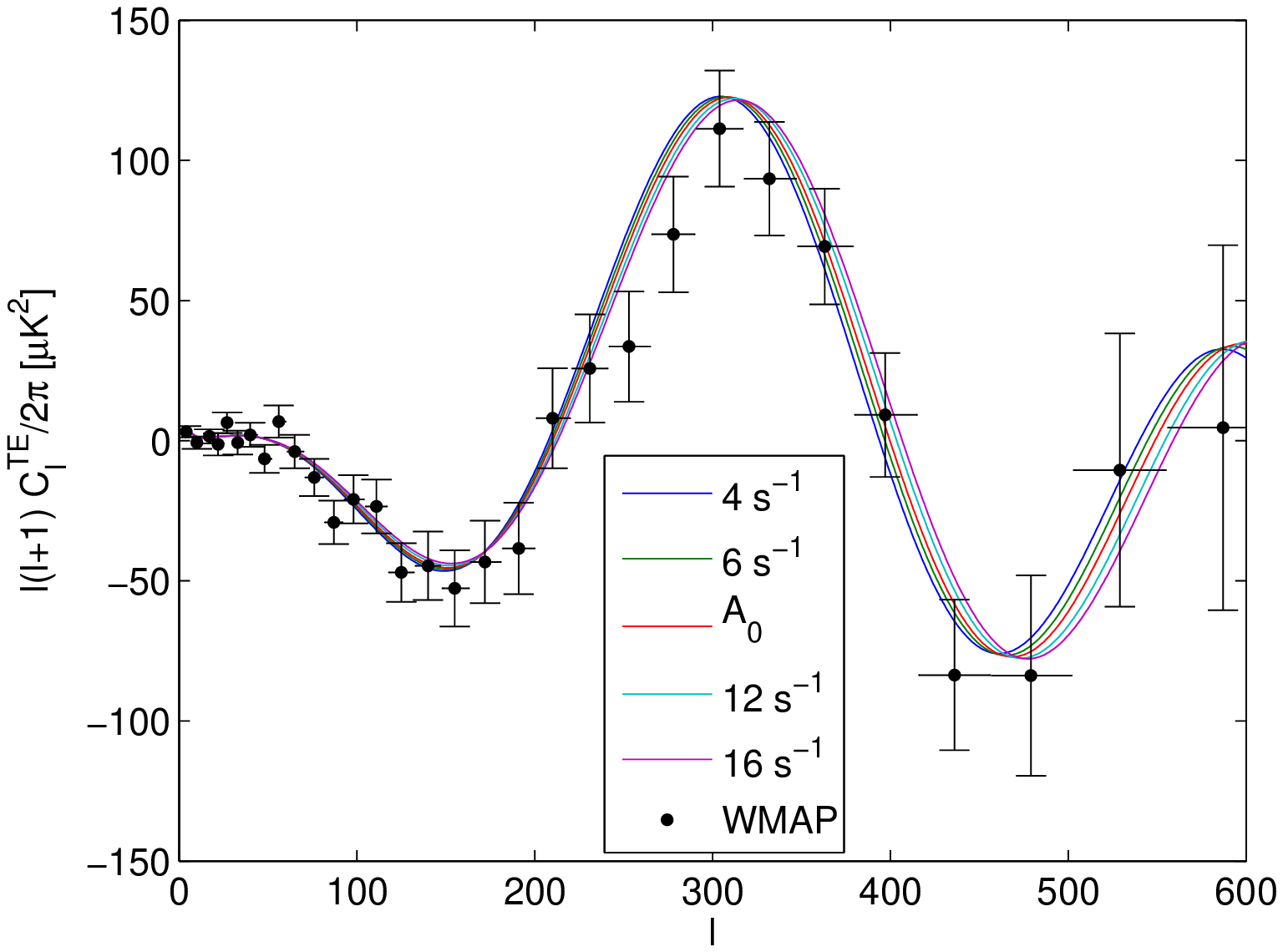}
\includegraphics[width=0.34\textheight]{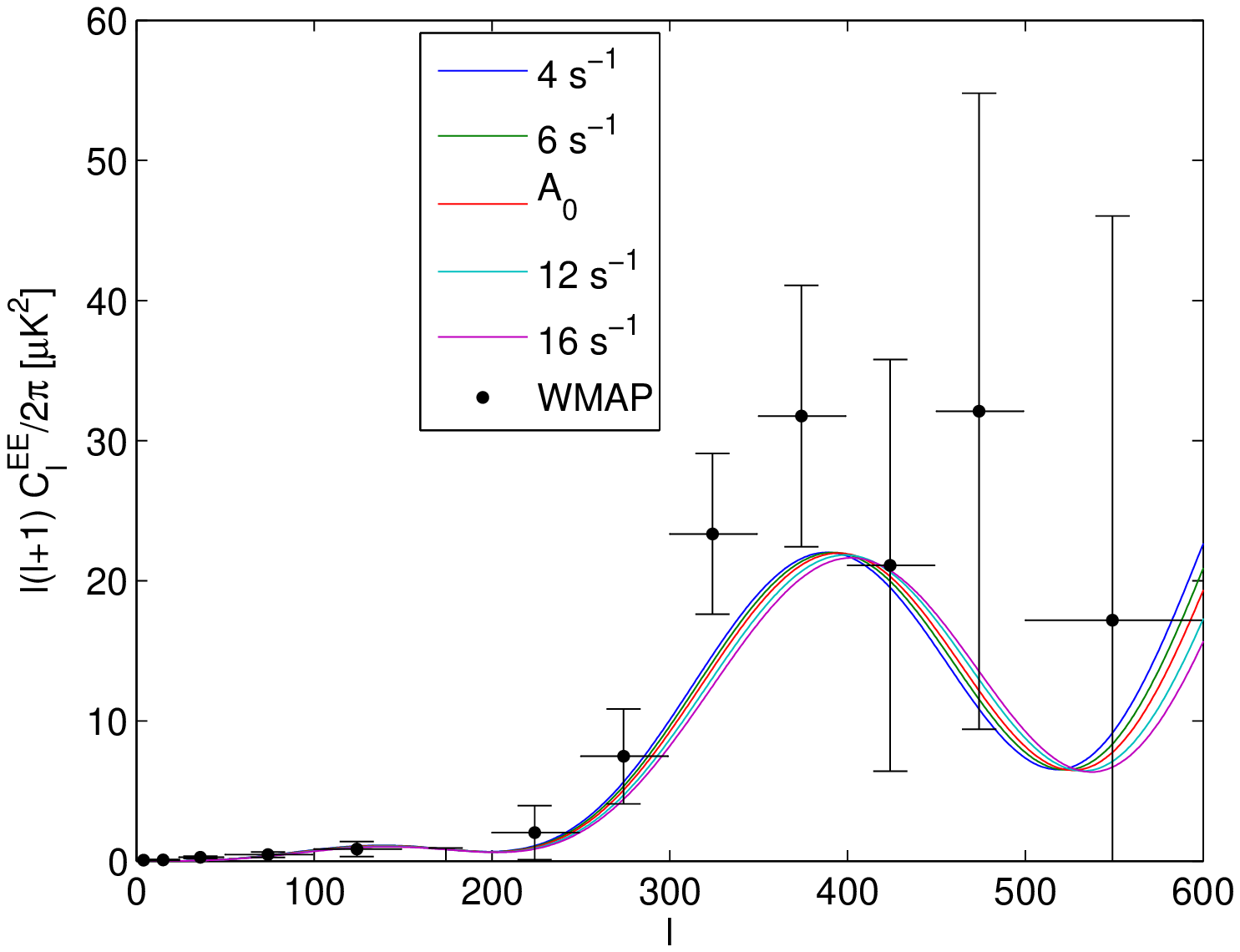}
\caption{CMB temperature power spectrum (top left), TE correlation (top right) and power spectrum of E mode polarization (bottom)
are plotted respectively for various values of
the decay rate $A_{2s1s}$ with $A_{0}$ being the theoretical prediction $%
8.22458\,\mathrm{s}^{-1}$. In the same plot, we show the binned WMAP data.}
\label{Cl}
\end{figure}
In case of the Gaussian fluctuations the decomposition coefficients satisfy
the following statistical properties %
\citep{Modern_Cosmology,Inflation,Foundations_Cosmology,Cosmology}: 
\begin{eqnarray}
\langle a_{T,lm}a_{T,l^{\prime }m^{\prime }}^{\ast }\rangle 
&=&C_{l}^{TT}\delta _{ll^{\prime }}\delta _{mm^{\prime }}, \\
\langle a_{E,lm}a_{E,l^{\prime }m^{\prime }}^{\ast }\rangle 
&=&C_{l}^{EE}\delta _{ll^{\prime }}\delta _{mm^{\prime }}, \\
\langle a_{T,lm}a_{E,l^{\prime }m^{\prime }}^{\ast }\rangle 
&=&C_{l}^{TE}\delta _{ll^{\prime }}\delta _{mm^{\prime }},
\end{eqnarray}%
where $\langle \ldots \rangle $ denotes the average over an ensemble of
universes. The power spectra for the temperature fluctuations $C_{l}^{TT}$,
for the E mode of polarization $C_{l}^{EE}$ and for the TE cross correlation 
$C_{l}^{TE}$, provide us invaluable information about early Universe 
\citep{Modern_Cosmology,Inflation,Foundations_Cosmology,Cosmology}.
Since the rate of cosmic recombination during its most important stage is
mainly determined by the two-photon decay rate $A_{2s1s}$  the correlation
functions above are rather sensitive to the particular numerical value of $%
A_{2s1s}$. 

By using \texttt{RECFAST} and \texttt{CAMB} code %
\citep{RECFAST1,RECFAST2,RECFAST3,CAMB} with small modifications, we have
computed CMB power spectra for various $A_{2s1s}$. In Fig. \ref{Cl} we show these spectra together with the WMAP data %
\citep{WMAP7:basic_result,WMAP5:powerspectra,WMAP7:powerspectra}. 
Though we show only the binned data
not to clutter the plots, we used the full WMAP data likelihood in the
analysis in the next section.
As noticed
in Fig. \ref{Cl}, the shape of $C_{l}^{TT}$, $C_{l}^{TE}$
and $C_{l}^{EE}$ are sensitive to the value of $A_{2s1s}$. 
As shown in Fig. \ref{xe}, the last scattering surface is affected by the variation of $A_{2s1s}$.
The acoustic peaks of temperature anisotropy is, in particular, sensitive to the shift of the last scattering surface, and polarization is affected by the change in the thickness of the last scattering surface.
Therefore, the EE powerspectrum and TE correlation as well as TT powerspectrum are essential to provide the tight contraint on the values of $A_{2s1s}$.
Additionally, we find that CMB anisotropy at high multipoles is affected more than those at low multipoles, which may be understood by the fact that the shift and thickness change of the last scattering surface is negligible in comparison with the physical scales of low multipoles.

\section{Constraints from the recent observational data}

\label{analysis} As discussed in the previous sections, the CMB power
spectra are sensitive to the value of the decay rate $A_{2s1s}$. Noting
this, we constrained the value $A_{2s1s}$ by the WMAP CMB data \citep{WMAP7:basic_result,WMAP7:powerspectra}. 
For a cosmological model, we assumed $\Lambda$CDM + SZ effect + weak-lensing.
Since the co-moving distance to the last scattering has dependence on Hubble expansion, there exist some level of parameter degeneracy between $A_{2s1s}$ and the Hubble parameter. 
From Fig. \ref{like2}, we may see some degeneracy with respect to $A_{2s1s}$, and Hubble parameter.
Besides WMAP CMB data, we additionally used data such as the Hubble Constant measurement with the Hubble Space Telescope (HST), Baryonic Acoustic
Oscillation (BAO) data from SDSS and WiggleZ, and Big Bang Nucleosynthesis
constraint \citep{HST,wigglez_bao,sdss_bao,BBN}. 
These data are not directly sensitive to the $A_{2s1s}$, but they enhance the constraint on $A_{2s1s}$ by reducing the uncertainty of Hubble parameter. 
We ran the \texttt{CosmoMC} with slight modifications on a MPI cluster with 6 chains \citep{CAMB,CosmoMC}. For the convergence
criterion, we adopted the Gelman and Rubin's \textquotedblleft variance of
chain means\textquotedblright\ and set the R-1 statistic to $0.03$ for
stopping criterion \citep{Gelman:inference,Gelman:R1}.
Analyzing the Markov chains produced by the \texttt{CosmoMC}, we obtained a
the best-fit values of the parameters and their confidence intervals. From the
analysis, we impose the following constraint on the decay rate $A_{2s1s}$: $%
 6.24\mathrm{s}^{-1}<A_{2s1s}< 11.89\mathrm{s}^{-1}$ and $ 4.47\mathrm{s}^{-1}<A_{2s1s}<14.67\mathrm{s}^{-1}$ with 1 and 2 $\sigma $ confidence
respectively. 
\begin{table}
\centering
\begin{tabular}{rrr}
\hline\hline
symbol & description & value \\ \hline
$\Omega_{b}\,h^2$ & baryonic density $\times h^2$ & $%
0.0223^{+0.0008}_{-0.0003}$ \\ 
$\Omega_{DM}\,h^2$ & cold dark matter density $\times h^2$ & $%
0.114^{+0.006}_{-0.005}$ \\ 
$\tau$ & optical depth & $0.081^{+0.021}_{-0.009}$ \\ 
$n_s$ & spectral index & $ 0.965^{+0.015}_{-0.013}$ \\ 
$\log[10^{10}A_s]$ & scalar amplitude & $ 3.08^{+0.04}_{-0.02}$ \\ 
$A_{\mathrm{sz}}$  & fitting coefficient of SZ effect &  $0.04^{+1.96}_{-0.04}$\\
$H_0$ [km/s/Mpc] & Hubble constant & $ 69.06^{+3.14}_{-0.96}$ \\ 
$A_{2s1s}$ [$\mathrm{s}^{-1}$] & two-photon decay rate & $%
 8.04^{+3.85}_{-1.8}$ \\ \hline
\end{tabular}%
\caption{the best-fit values of cosmological parameters $+$ $A_{2s1s}$ with 1%
$\protect\sigma $ interval indicated. The scalar amplitude $A_{s}$ is at the 
$k_{0}=0.05\,[\mathrm{Mpc^{-1}}]$.}
\label{bestfit}
\end{table}
In Table \ref{bestfit}, we show the best-fit values of the
decay rate $A_{2s1s}$ and cosmological parameters with 1 $\protect\sigma $ interval indicated. 
\begin{figure}
\centering
\includegraphics[scale=.93]{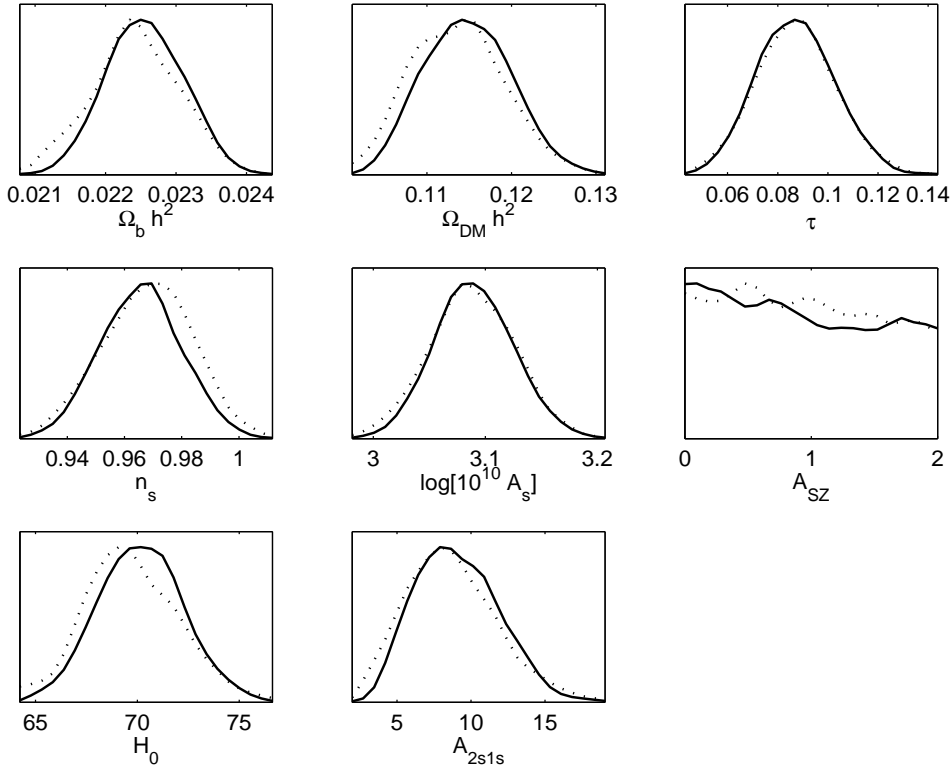}
\caption{Likelihood of $A_{2s1s}$: a solid lines denote a marginalized
likelihood and a dotted line a mean likelihood (refer to \citep{CosmoMC} for
distinction between them).}
\label{like1}
\end{figure}
\begin{figure}
\centering
\includegraphics[scale=.93]{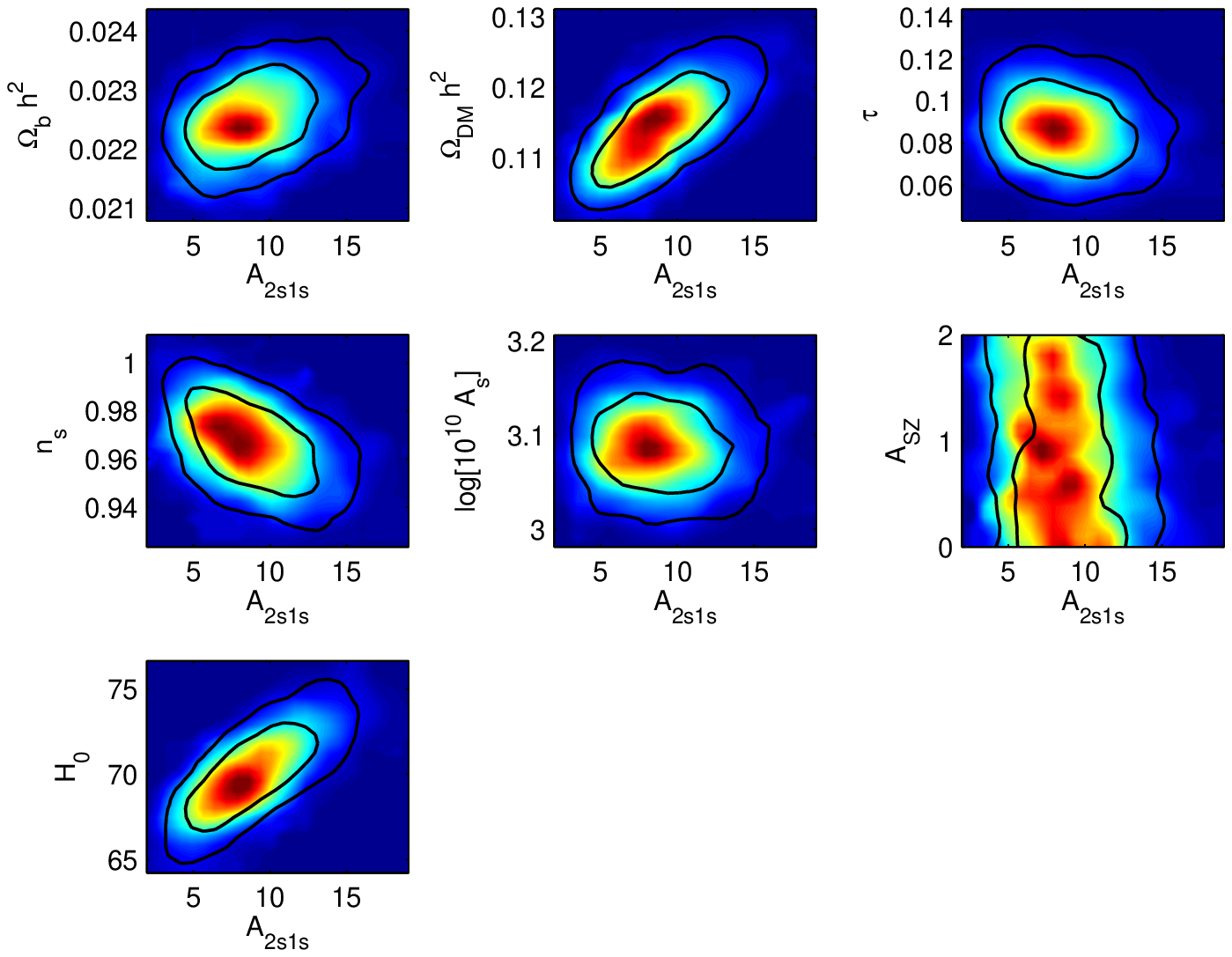}
\caption{The marginalized likelihood in the plane of $A_{2s1s}$ versus
others parameters. The redder(bluer) color denotes higher(lower) likelihood.
Two contour lines correspond to 1$\protect\sigma $ and 2$ %
\protect\sigma $ levels respectively.}
\label{like2}
\end{figure}
In Fig. \ref{like1}, we show the likelihood distribution for each parameter
and in Fig. \ref{like2} the marginalized likelihoods in the plane of $%
A_{2s1s}$ versus other parameters. From Fig. \ref{like2}, we infer some level of parameter degeneracy between Hubble constant and $A_{2s1s}$.
Fig. \ref{like2} also shows slight parameter degeneracy with the spectral index $n_s$. However, the spectral index, which determines the shape of the primordial power spectrum,
often have slight level of degeneracy with other cosmological parameters, since the variation of the spectral index can mimic the variation of other parameters more or less. 
The degeneracy with $\Omega_{\mathrm{DM}}\,h^2$ seen in Fig. \ref{like2} is attributed to the factor $h^2$.

\begin{table}[!hb]
\centering
\begin{tabular}{ccc}
\hline\hline
Beam (FWHM) [arcminute]  & temperature noise [$\mu \mathrm{K}$] & polarization noise [$\mu \mathrm{K}$] \\ \hline
9.5 & 6.8 & 10.9 \\
7.1 & 6.0 & 11.4 \\
5 & 13.1 & 26.7 \\ \hline
\end{tabular}%
\caption{Assumed instrumental properties of the PLANCK mock data.}
\label{futurcmb}
\end{table}
As discussed previously, CMB polarization is sensitive to the $A_{2s1s}$, and CMB anisotropy on smaller angular scales is more sensitive than those on large scales.
Therefore, the upcoming Planck data will provides a very tight constraint on $A_{2s1s}$, thanks to the low noise polarization data and the temperature data of high angular resolution.
In order to assess the constraining power of the PLANCK surveyor
data, we made the parameter forecast, using the PLANCK mock data. The PLANCK
mock data was generated up to the multipole $l=2000$ by the publically
available \texttt{FUTURCMB} code with the expected sensitivity of the PLANCK
surveyor \citep{FUTURCMB,Planck_bluebook}, where we assumed the WMAP concordance model and the decay rate $A_{2s1s}$ to $8.22458\,\mathrm{s}^{-1}$. 
For the mock data, we assumed three channels with a sky fraction $0.65$.
The assumed instrumental properties of the three channels are summarized in Table \ref{futurcmb}.
For the mock data constraint, we did not use the lensing convergence power spectrum, but only TT, TE, EE power spectrum.
From the run of the \texttt{CosmoMC}
with the mock data, we found the estimation error on $A_{2s1s}$ is $0.486\;%
\mathrm{s}^{-1}$, which is less than $6\,\%$ of the central value. To be
specific, the constraints imposed by the PLANCK mock data are $8.086\mathrm{s%
}^{-1}<A_{2s1s}<9.037\mathrm{s}^{-1}$ and $7.613\mathrm{s}%
^{-1}<A_{2s1s}<9.505\mathrm{s}^{-1}$ at 1$\sigma$ and 2$\sigma$ level respectively.
The improvement mainly comes from temperature data on small angular scales and low noise polarization data, which are sensitive to the value of $A_{2s1s}$.
We may further enhance the constraint by adding non-CMB data to PLANCK mock data. However, the improvement by non-CMB data mainly arise from the tightened constraint on Hubble parameter, which is already well constrained by PLANCK mock data.

\section{Discussion}

\label{Discussion} We have shown that the recent WMAP TT and TE data sets in
combination with the BAO and HST data allow us to constrain the range
of uncertainties of the decay rate $A_{2s1s}$ within the interval $+3.85,-1.8$%
, which corresponds in average to $\pm 34\%$. The PLANCK mock data up to the
multipole $l=2000$ with the expected sensitivity of the PLANCK surveyor can
significantly reduce the level of error bars down to $8.086\mathrm{s}%
^{-1}<A_{2s1s}<9.037\mathrm{s}^{-1}$, around the most probable value $%
A_{2s1s}\simeq 8.2\,\mathrm{s}^{-1}$. This estimation clearly illustrate
that the theory of recombination, based on the theoretical value of $A_{2s1s}
$, is self -consistent. Actually, our analysis confirm prediction made in %
\citep{Ionization_history_DM}, that any modifications of the kinetic of
recombination, which could change the fraction of ionization at redshifts $%
800\leq z\leq 1100$ by factor $\Delta x_{e}/x_{e}=\delta $ would lead to
corresponding changes of the TT and TE power spectrum $\delta
C(l)/C(l)\simeq \delta $. Taking into account that the natural limit of
uncertainties in the power spectrum is the cosmic variance, one can get ; $%
\delta C(l)/C(l)\simeq (l+0.5)^{-\frac{1}{2}}\sim \delta $. 
In the model discussed above, the uncertainties of the decay rate $A_{2s1s}$
are in order of $\Delta A/A\simeq 6\%$ leading to $\delta \sim 0.5\Delta A/A$
(see Fig.1). Thus, for $\bar{l}=2000$ the corresponding constraint is given
by $\Delta A/A\simeq 1/\sqrt{\bar{l}}\sim 4\%$, which is close to the
constrain, given by CosmoMC approach. Thus, if the systematic effects for
the forthcoming PLANCK data release would be comparable to the cosmic variance limit
for the range of multipoles \ around $\bar{l}=2000$, our prediction of
uncertainties of the decay rate $A_{2s1s}$ would have experimental
confirmation.

\acknowledgments
We thank an anonymous referee for very helpful comments, which leads to important improvements in the clarity of this paper.
We are grateful to Theodor W. H\"{a}nsch, Subir Sarkar and Rashid Sunyaev for helpful discussions.
We acknowledge the use of the Legacy Archive for Microwave Background Data
Analysis (LAMBDA). Our data analysis made the use of the \texttt{CosmoMC}
package and \texttt{FUTURCMB} code \citep{CAMB,CosmoMC,FUTURCMB}. This work
is supported in part by Danmarks Grundforskningsfond, which allowed the
establishment of the Danish Discovery Center. CB and TT acknowledges partial
support by ASI through ASI/INAF Agreement I/072/09/0 for the Planck LFI
Activity of Phase E2 and by MIUR through PRIN 2009. The work of VM was
supported by \textquotedblleft Chaire Internationale de Recherche Blaise
Pascal financ\'{e}e par l'Etat et la R\'{e}gion d'Ile-de-France, g\'{e}r\'{e}%
e par la Fondation de l'Ecole Normale Sup\'{e}rieure\textquotedblright , by
TRR 33 \textquotedblleft The Dark Universe\textquotedblright\ and the
Cluster of Excellence EXC 153 \textquotedblleft Origin and Structure of the
Universe\textquotedblright .

\bibliographystyle{JHEP}
\bibliography{/home/tac/jkim/Documents/bibliography}
\end{document}